\documentclass[prd, superscriptaddress, twocolumn, amsfonts, amssymb, amsmath, showpacs]{revtex4-2}
\usepackage{bm}
\usepackage{amsfonts}
\usepackage{latexsym}
\usepackage[latin1]{inputenc}
\usepackage{graphicx}
\usepackage{amsmath}
\usepackage{palatino}
\usepackage{mathpazo}
\usepackage[british]{babel}
\usepackage{hhline}
\usepackage{multirow}
\usepackage{textcomp}
\usepackage{float}
\usepackage{booktabs}
\usepackage{dcolumn}
\usepackage{hhline}
\usepackage{multirow}
\usepackage{ragged2e}
\usepackage{hyperref}
\hypersetup{colorlinks,citecolor=blue}
\hypersetup{colorlinks=true,linkcolor=red,filecolor=magenta,    urlcolor=cyan}
\usepackage{amsmath}
\usepackage{xcolor}
\usepackage{orcidlink}
\usepackage{epsfig}
\usepackage{caption}
\usepackage{subcaption}
\usepackage{commath}
\def\jnl@style{\it}
\def\aaref@jnl#1{{\jnl@style#1}}

\def\aaref@jnl#1{{\jnl@style#1}}

\def\aj{\aaref@jnl{AJ}}                   
\def\apj{\aaref@jnl{ApJ}}                 
\def\apjl{\aaref@jnl{ApJ}}                
\def\apjs{\aaref@jnl{ApJS}}               
\def\apss{\aaref@jnl{Ap\&SS}}             
\def\aap{\aaref@jnl{A\&A}}                
\def\aapr{\aaref@jnl{A\&A~Rev.}}          
\def\aaps{\aaref@jnl{A\&AS}}              
\def\mnras{\aaref@jnl{Mon.~Not.~Roy.~Astron.~Soc.}}             
\def\prd{\aaref@jnl{Phys.~Rev.~D}}        
\def\prc{\aaref@jnl{Phys.~Rev.~C}}  
\def\prl{\aaref@jnl{Phys.~Rev.~Lett.}}    
\def\qjras{\aaref@jnl{QJRAS}}             
\def\skytel{\aaref@jnl{S\&T}}             
\def\ssr{\aaref@jnl{Space~Sci.~Rev.}}     
\def\zap{\aaref@jnl{ZAp}}                 
\def\nat{\aaref@jnl{Nature}}              
\def\aplett{\aaref@jnl{Astrophys.~Lett.}} 
\def\apspr{\aaref@jnl{Astrophys.~Space~Phys.~Res.}} 
\def\physrep{\aaref@jnl{Phys.~Rep.}}      
\def\physscr{\aaref@jnl{Phys.~Scr}}       
\def\commat{\aaref@jnl{Comm.~Math.~Phys.}}              
\def\science{\aaref@jnl{Science}}               
\def\cqg{\aaref@jnl{Classical Quant.~Grav.}}            
\def\jpcs{\aaref@jnl{JPCS}}                                     
\def\ijmpd{\aaref@jnl{Int.~J.~Mod.~Phys.~D}}                    
\def\grg{\aaref@jnl{Gen.~Relat.~Gravit.}}               
\def\rpp{\aaref@jnl{Rep.~Prog.~Phys.}}          
\def\npa{\aaref@jnl{Nucl.~Phys.~A}}        
\def\lrr{\aaref@jnl{Living Rev.~Rel.}}                   
\def\jcap{\aaref@jnl{J.~Cosmology Astropart.~Phys.}}    
\def\rmp{\aaref@jnl{Rev.~Mod.~Phys.}}   
\def\epjc{\aaref@jnl{Eur.~Phys.~J.~C}} 
\def\plb{\aaref@jnl{~Phy.~Lett.~B}} 
\def\mpla{\aaref@jnl{Mod.~Phy.~Lett.~A}} 
\def\arxiv{\aaref@jnl{arxiv.org}}

\allowdisplaybreaks[1]

\begin{document}
\color{black}       
%
\title{\bf Attractor behaviour of $f(T)$ modified gravity and the cosmic acceleration}

\author{L.K. Duchaniya \orcidlink{0000-0001-6457-2225}}
\email{duchaniya98@gmail.com}
\affiliation{Department of Mathematics, Birla Institute of Technology and Science-Pilani, Hyderabad Campus,  Hyderabad-500078, India.}

\author{Kanika Gandhi \orcidlink{0000-0002-5647-2088}}
\email{kanika.00@hotmail.com}
\affiliation{Department of Physics,
Birla Institute of Technology and Science-Pilani, Hyderabad Campus,
Hyderabad-500078, India.}

\author{B. Mishra\orcidlink{0000-0001-5527-3565}}
\email{bivu@hyderabad.bits-pilani.ac.in}
\affiliation{Department of Mathematics, Birla Institute of Technology and Science-Pilani, Hyderabad Campus, Hyderabad-500078, India.}

\begin{abstract}
{\bf{Abstract}:} In this paper, we have performed the dynamical system analysis of $f(T)$ gravity cosmological models at both background and perturbation levels. We have presented three models pertaining to three distinct functional forms of $f(T)$. The first form is that of the logarithmic form of the torsion scalar $T$, the second one is in the power law form, and the third one is the combination of the first two forms. For all these three forms of $f(T)$, we have derived the corresponding cosmological parameters in terms of the dynamical variables. Subsequently, the critical points are obtained and the condition(s) of its existence has been derived. Critical points of each model have been analysed individually and the corresponding cosmology are derived. The stability behaviour of these critical points are discussed from the behaviour of the eigenvalues and the phase portraits. At least one stable node has been obtained in each of these models. Further from the evolution plots of the cosmological parameters, the accelerating behaviour of the cosmological models are also verified.    
\end{abstract}

\maketitle

\section{Introduction} \label{SEC-I}
In recent times, several cosmological models are being presented on the modified or extended theories of gravity to address the issue of late-time cosmic phenomena. The extension or modification are basically on the Einstein-Hilbert action of General Relativity (GR), though GR has been successful in addressing many complex issues of the Universe. However, some issues are yet to be resolved, hence the modification has been inevitable. For example, the most recent findings of the cosmological research on the accelerated expansion of the Universe \cite{Riess:1998cb, Perlmutter:1998np, Tegmark_2004a, Spergel_2003, Riess_2004}. This behaviour is on the claim of the presence of some exotic form of energy, termed Dark Energy (DE).  Theoretically, through the cosmological constant, $\Lambda$CDM (Cold Dark Matter) describes the DE, however, it needs to be fine-tuned in spite of its observational successes  \cite{Peebles:2002gy, Padmanabhan_2003, RevModPhys.61.1}. One can include the matter field such as the Canonical scalar field, vector field, phantom field etc. in the dark energy sector to frame the cosmological model \cite{COPELAND_2006, Cai_2010}. Else the gravitational sector can also be modified, resulting in the modified or extended gravity \cite{Capozziello:2011et, Nojiri_2011a}. One such modification in the gravitational sector is the 
teleparallel equivalent of GR (TEGR) \cite{Einstein28, Maluf:1994j, Aldrovandi:2013wha, Pereira:2019woq}. In this framework, unlike in GR, where the Levi-Civita connection denotes curvature, in TEGR the Weitzenb$\ddot{o}$ck connection signifies torsion in teleparallelism \cite{Weitzenbock1923}. The tangent space at each point in space-time is based on four linearly independent tetrad fields. In this approach, the modified theory can be formulated which leads to the second-order equations in four-dimensional space-time. The first derivative of tetrad leads to the derivation of torsion tensor. One of the simple modification of TEGR is the $f(T)$ gravity, where the matter Lagrangian is an arbitrary function of the torsion scalar $T$ \cite{Ferraro:2006jd, Bengochea:2008gz, Tamanini:2012hg, Cai:2015emx}. Some other modifications of TEGR are:  $f(T,B)$ gravity \cite{Bahamonde:2015zma, Escamilla-Rivera:2019ulu}, $B$ being the boundary term that represents the difference between the Ricci and torsion scalar; $f(T,\mathcal{T})$ \cite{Harko_2014a, Duchaniya_2024pd} gravity, where $\mathcal{T}$ represents the trace of the energy-momentum tensor; $f(T, T_{\mathcal{G}})$ gravity \cite{Kofinas:2014owa, Lohakare_2023a}, where $\mathcal{G}$ denotes the  Gauss-Bonnet term; and $f(T,\phi)$ \cite{PhysRevD.97.104011, Duchaniya_2023}, $\phi$ represents the scalar field and so on. Here onwards, we abbreviate Teleparallel Gravity as TG.\\

We shall discuss some of the recent research on $f(T)$ gravity available in the literature pertaining to the issue of late-time cosmic expansion. The evolutionary behaviour guided from the equation of state (EoS) parameter in the exponential and logarithmic setting of $f(T)$ has been shown in Ref. \cite{Bamba_2011}. Bamba et. al \cite{Bamba_2013bc} have examined some conformal issues of pure and extended teleparallel gravity also, they have proposed conformal scalar and gauge field theories and constructed conformal torsion gravity. Farrugia and Jackson \cite{Farrugia_2016a} have analysed the small perturbation $\delta$ about the Hubble parameter, and the matter-energy density $\delta_{m}$, to show the stability of flat FLRW metric in $f(T)$ gravity. In Ref. \cite{Anagnostopoulos_2019a}, the $f(T)$ gravity has been constrained at both background and perturbation levels with several data sources such as the Pantheon supernovae sample, Hubble constant measurements,  cosmic microwave shift parameter, redshift-space distortion measurement. Zhao et al. \cite{Zhao_2022} have calculated the quasinormal modes frequencies of a test massless scalar field around static black holes solutions in $f(T)$ gravity. The second-order primordial scalar-induced gravitational waves produced by primordial black hole Poisson fluctuations in the context of $f(T)$ modified gravity studied by Papanikolaou et al.\cite{Papanikolaou_2023}. Jackson et al. \cite{Said_2020a} have investigated that the violation of the distance-duality relation is directly linked with a temporal variation of the electromagnetic fine-structure constant with different forms of $f(T)$ gravity models. Paliathanasis et al. \cite{Paliathanasis_2016} have studied the cosmological scenario in $f(T)$ gravity and they have found analytical solutions for an isotropic and homogeneous universe containing a dust fluid and radiation and for an empty anisotropic Bianchi type I Universe.   
 The existence, and study stability, of the Kasner vacuum solution of Bianchi type I for the modified $f(T)$ teleparallel gravity studied in Ref.\cite{Paliathanasis_2018}. Agostino et. al \cite{D_Agostino_2018} have investigated the growth rate of matter perturbations in the context of teleparallel dark energy in a flat universe and also, investigated the dynamics of different theoretical scenarios based on specific forms of the scalar field potential. Capozziello et al. \cite{Capozziello_2019a}  have discussed the cosmographic method and its applications to cosmological models derived from extended or modified theories of gravity. Some more research on $f(T)$ gravity based late time cosmic expansion issue can be seen in  $f(T)$ gravity \cite{Dent_2011a, Cai:2015emx, Duchaniya_2022, Briffa_2022a}.  Also, Coley et al. \cite{Coley_2307_12930} studied a class of teleparallel geometries that have a seven-dimensional group of affine symmetries, which is a subgroup of the Killing symmetries of the de Sitter metric. These geometries are known teleparallel de Sitter (TdS) geometries and are analogues of the de Sitter geometry in GR. A set of six-dimensional Robertson-Walker affine symmetries with invariant co-frames and spin connections is displayed and discussed in Ref.\cite{Coley_2310_14378}. \\

While framing the cosmological models with modified gravity theories, normally we come across several challenging equations and ambiguous initial conditions. As a result, it has become extremely difficult to obtain the analytical solution and therefore to obtain at least the qualitative behaviour, the dynamical system analysis can be employed \cite{wainwrightellis1997, Basilakos_2019a}. It is well known that the general cosmological system can be interpreted in several ways, but the asymptotic behaviour must converge over time. Cosmological equations can be linked to stable critical points. The evolutionary phases of the Universe such as the matter and radiation phases are also represented by the critical points of the autonomous system and note these critical points are unstable or saddle nodes. In the context of modified gravity cosmological models, some of the recent research adopted the dynamical system techniques \cite{Dutta_2018az, Awad_2018, Narawade_2022a, Duchaniya_2023, Kadam:2022lgq, Franco2021aca, Khyllep2023, Agrawal_2023}. Moreover, we discuss here some of the recent work on the dynamical system analysis in $f(T)$ gravity.  At the background level, Wu and Yu \cite{Wu_2010a} have carried out the power-law form of $f(T)$ to study the dynamical system analysis and obtained a stable de Sitter phase along with an unstable radiation-dominated and an unstable matter-dominated phase. Hohmann et al. \cite{Hohmann_2017a}  have derived a two-dimensional dynamical system from the flat FLRW cosmological field equations of a generic $f(T)$ gravity theory. Mirza and Oboudiat \cite{Mirza_2017a}  have investigated the cosmological solutions of the $f(T)$ gravity theory using the method of dynamical systems at the background level. Inspired by the most latest studies on dynamical system analysis by Duchaniya et al.\cite{Duchaniya_2022} at the background level. In this article, we explore the dynamical system analysis at both background and perturbation level in $f(T)$ gravity. The structure of the article is as follows:  We briefly describe the foundations of teleparallel and the operation of the $f(T)$ gravity and its field equations in section-\ref{SEC-II}. The phase space analysis of the three models governed by the form of the function $f(T)$ has been performed in section-\ref{SEC-III}. Finally the conclusions are described in section-\ref{SEC-IV}.\\ 

\section {$f(T)$ gravity field equations} \label{SEC-II}
In teleparallel framework, tetrads may be employed as the dynamical variable instead of the metric tensor. The tetrad $e^{A}_{\mu}$  must fulfill the orthogonality conditions, i.e.
\begin{equation}\label{1A}
 e^{A}_{\mu} e^{\mu}_{B}= \delta^{A}_{B}, \hspace{3cm} e^{\mu}_{A} e^{A}_{\nu}= \delta^{\mu}_{\nu}\,.
\end{equation}

The Latin index runs over $0, 1, 2, 3$ denote the tangent space of the manifold, whereas the Greek index runs over $0, 1, 2, 3$ is the coordinate of space-time on the manifold and $e^{\mu}_{A}$  represents the inverse tetrad. These tetrads relate to the metric tensor $g_{\mu \nu}$ through the Minkowski space-time as,
\begin{equation} \label{1}
 g_{\mu \nu}=\eta_{AB} e_{\mu}^{A} e_{\nu}^{B}.    
\end{equation}
The Minkowski metric $\eta_{AB}=diag(1,-1,-1,-1)$. The TG and the Weitzenb$\ddot{o}$ck connection \cite{Weitzenbock1923} linked as, 
\begin{equation}\label{1B}
\Gamma^{\lambda}_{\nu \mu}\equiv e^{\lambda}_{A} \partial_{\mu} e^{A}_{\nu}.    
\end{equation}
In the TG framework, the torsion tensor is an antisymmetric part of Weitzenb$\ddot{o}$ck connection, defined as
\begin{equation}\label{2}
T^{\lambda}_{\mu \nu}\equiv\hat\Gamma^{\lambda}_{\nu \mu}-\hat\Gamma^{\lambda}_{\mu \nu}=e^{\lambda}_{A} \partial_{\mu} e^{A}_{\nu}-e^{\lambda}_{A} \partial_{\nu} e^{A}_{\mu}.
\end{equation}
The superpotential tensor can be defined as,
\begin{equation}\label{1C}
S_{\rho}^{~~\mu \nu}\equiv\frac{1}{2}(K^{\mu \nu}_{~~~\rho}+\delta^{\mu}_{\rho}T^{\alpha \nu}_{~~~\alpha}-\delta^{\nu}_{\rho}T^{\alpha \mu}_{~~~\alpha}),    
\end{equation}
where the contortion tensor,
\begin{equation}\label{1D}
 K^{\mu \nu}_{~~~\rho}\equiv \frac{1}{2}(T^{\nu \mu}_{~~~\rho}+T_{\rho}^{~~\mu \nu}-T^{\mu \nu}_{~~~\rho}).   
\end{equation}
Further from the contraction of the torsion tensor, the torsion scalar can be obtained as,
\begin{equation}\label{3}
T \equiv \frac{1}{4} T^{\rho \mu \nu} T_{\rho \mu \nu}+\frac{1}{2} T^{\rho \mu \nu} T_{\nu \mu \rho}-T_{\rho \mu}^{~~\rho} T^{\nu \mu}_{~~\nu}\,.
\end{equation}
The $f(T)$ gravity is to generalize $T$ to a function $T+f(T)$, whose action \cite{Cai:2015emx} can be defined as,  
\begin{equation}\label{4}
S = \frac{1}{16 \pi G}\int d^{4}xe[T+f(T)+\mathcal{L}_{m}],
\end{equation}

where $G$ is the gravitational constant and $e=det[e^A_{\mu}]=\sqrt{-g}$ and $\mathcal{L}_{m}$ be the total matter Lagrangian. We consider the natural system, $\kappa ^2=8\pi G=c =1$. Varying the action Eqn. \eqref{4} with respect to the vierbein, the gravitational field equations can be obtained as,
\begin{eqnarray}\label{5}
&&e^{-1}\partial_{\mu}(e e^{\rho}_{A}S_{\rho}^{~\mu \nu})[1+f_{T}] +e^{\rho}_{A}S_{\rho}^{~\mu \nu}\partial_{\mu}(T)f_{TT}+ \frac{1}{4} e^{\nu}_{A} T \nonumber \\ &&-e^{\lambda}_{A}T^{\rho}_{~\mu \lambda}S_{\rho}^{~ \nu\mu}[1+f_{T}] +\frac{1}{4}e^{\nu}_{A} f(T)=4 \pi G e^{\rho}_{A}T_{~\rho}^{~~\nu}.
\end{eqnarray}

We denote $f=f(T)$ and the first and second derivative of $f$ with respect to $T$ as $f_{T}$ and $f_{TT}$ respectively and  $ T_{~\rho}^{~~\nu} $ be the total matter energy-momentum tensor. For the cosmological scenario, we consider the Friedmann-Lemaitre-Robertson-Walker(FLRW) space-time as,
\begin{equation}
ds^{2}=dt^{2}-a^{2}(t)[dx^{2}+dy^{2}+dz^{2}], \label{6}
\end{equation}

with $a(t)$ be the scale factor and  $ e^{A}_{\mu}\equiv diag(1,a(t),a(t),a(t))$. The field equations of $f(T)$ gravity for FLRW space-time are,  
\begin{equation}
3H^2 = 8\pi G \rho_m-\frac{f}{2}+Tf_T \label{7},
\end{equation}
\begin{equation}
\dot{H} = -\frac{4\pi G(\rho_{m}+p_{m})}{1+f_T+2Tf_{TT}}. \label{8} 
\end{equation}

The Hubble parameter $H\equiv\frac{\dot{a}}{a}$ with an over dot denotes the derivative with respect to cosmic time $t$. The matter-energy density and pressure respectively $\rho_{m}$ and $p_{m}$ and the total energy-momentum tensor is comprised of the matter sector. Now the field equations of $f(T)$ gravity in the dark energy sector pressure ($p_{de}$) and energy density($\rho_{de}$) can be defined as,
\begin{equation}
    \rho_{de} \equiv \frac{1}{16 \pi G}\left[-f+2Tf_{T}\right], \label{9}
\end{equation}
\begin{equation}
    p_{de} \equiv -\frac{1}{16 \pi G}\left[\frac{-f+Tf_{T}-2T^{2}f_{TT}}{1+f_{T}+2Tf_{TT}}\right].\label{10}
\end{equation}
Substituting the torsion scalar $T=-6H^2$, the EoS parameter of the dark energy sector ($\omega_{de}$) can be obtained as, 
\begin{equation}\label{11}
\omega_{de}=-1+\frac{\left(f_T+2Tf_{TT}\right)\left(-f+T+2Tf_T\right)}{(1+f_{T}+2Tf_{TT})(-f+2Tf_{T})}.
\end{equation}
Further, the total EoS ($\omega_{tot}$) and deceleration parameter ($q$) can be defined as,
\begin{eqnarray}
\omega_{tot.}&=&-1-\frac{2\dot{H}}{3H^{2}}\equiv \frac{p_m+p_{de}}{\rho_{m}+\rho_{de}}, \label{1E}\\  
q&=&-1-\frac{\dot{H}}{H^{2}}.\label{1F}
\end{eqnarray}
From Eqn. \eqref{7}, the constraint equation can be written in the form density parameters as,
\begin{equation} \label{15}
 \Omega_{de}+ \Omega_{m}=1\,.  
\end{equation}

We are intending to study the interacting cosmology by performing the dynamical system analysis. From the expressions of Eqn.\eqref{9} -- Eqn.\eqref{11}, we can see that the functional form of $f(T)$ is required for further study. Therefore we shall consider three distinct forms of $f(T)$ in the following section. 
\section{Phase space analysis} \label{SEC-III}
Here, we shall set up the dynamical system of the background and perturbed equations. In one of our previous works (Ref.\cite{Duchaniya_2022}), we studied dynamical system analysis at the background level. In this article, we are interested to broaden it further by including the impact of perturbations. To do this, the equation governing the growth of matter perturbations on sub-horizon scales can be invoked in the form \cite{Gannouji_2009a, Anagnostopoulos_2019a}  \\
\begin{equation}\label{23}
 \Ddot{\delta}_m+2 H \dot{\delta}_{m}=4 \pi G_{eff} \rho_{m} \delta_{m} \,, 
\end{equation} 

where $\delta_{m}=\frac{\delta \rho_{m}}{\rho_{m}}$ is the matter over density and the effective Newton's constant $G_{eff}(a)=G Y(a)$, with $G$ being the gravitational constant. Usually, $G_{eff}(a)$ is changeable, but the shape of $Y(a)$ set from the basic theory of gravity. Here, we keep the general perturbation method for $f(T)$ cosmology, if we have the form of $G_{eff}(a)$ or $Y(a)$ for the $f(T)$ gravity. So, we take the form of $Y(a)$ as in Ref. \cite{Zheng_2011},
\begin{equation}\label{24}
Y(a)= \frac{G_{eff}(a)}{G} =\frac{1}{1+f_{T}}\,,   
\end{equation}
From Eqn. \eqref{23}-- Eqn.\eqref{24}, we have the following 
\begin{equation}\label{25}
  \Ddot{\delta}_{m}+2 H \dot{\delta}_{m}= \frac{4 \pi G \rho_{m} \delta_{m}}{1+f_{T}}  
\end{equation}

Referring Eqn. \eqref{7}, Eqn.\eqref{8} and Eqn.\eqref{25}, initially we set up the dynamical variables of the background and perturbed equations for a general function of $f(T)$ as, 
\begin{equation} \label{26}
   x = -\frac{f}{6H^{2}},
    \hspace{1cm}
    y  = -2f_{T},
    \hspace{1cm}
    \sigma = \frac{d (ln \delta)}{d (lna)}\,. 
\end{equation}

Here, the variables $x$ and $y$ are related to describe the background evolution of the Universe whereas $\sigma$ quantifies the expansion of matter perturbations. So, when the matter density contrast is positive($\sigma>0$), means the matter perturbations are getting bigger, whereas for ($\sigma<0$), it gets smaller. We write Eqn. \eqref{7} in term of dynamical variables as,
\begin{equation}\label{27}
 x+y+\Omega_{m}=1\,.   
\end{equation}

To note here, the dynamical variables $x$  and $y$ define the dark energy density parameter ($\Omega_{de}$). Also, Eqn. \eqref{8}, in terms of dynamical variables can be,
\begin{equation}\label{28}
\frac{\dot{H}}{H^{2}} =  \frac{3(x+y-1)}{2(1+f_{T}+2Tf_{TT})}\,. 
\end{equation}

Using $x$, $y$, and $\sigma$ as phase space variables, we can perform $3D$ dimensional phase space analysis. In terms of the dynamical variables [Eqn. \eqref{26}], the cosmological equations can be written as an autonomous system as below,
\begin{eqnarray}
 \frac{dx}{dN}&=& -\frac{\dot{H}}{H^{2}}(y+2x)\,, \label{29} \\ 
 \frac{dy}{dN}&=& -4 \frac{\dot{H}}{H^{2}}(T f_{TT})\,, \label{30}\\
 \frac{d\sigma}{dN}&=& -\sigma(\sigma+2)-\frac{3(x+y-1)}{(2-y)}-\frac{\dot{H}}{H^{2}} \sigma \,,\label{31}
\end{eqnarray}

where $N=lna$. The EoS parameters and the deceleration parameter in terms of dimensionless variables are,
\begin{eqnarray}
 \omega_{de}&=&\frac{-2x-y+4Tf_{TT}}{2(x+y)(1+f_{T}+2Tf_{TT})} \,,\label{32} \\
\omega_{tot}&=& -1-\frac{(x+y-1)}{(1+f_{T}+2Tf_{TT})} \,, \label{33}  \\
q&=&-1-\frac{3(x+y-1)}{2(1+f_{T}+2Tf_{TT})} \,.  \label{34}
\end{eqnarray}

The critical points of the system Eqn. \eqref{29}--Eqn. \eqref{31} will be obtained to determine the dynamic growth of the system and the stability of these critical points will be examined. From the physical point of view, it is well known that the stable point ($\sigma>0$) implies continuous growth of matter perturbations and also we can say that system is not stable with respect to matter perturbation. But, a stable point having $\sigma<0$ denotes the reduction in matter perturbation. When $\sigma=0$ is at a stable point, it is considered that the changes in matter perturbation is always the same. To solve the dynamical system Eqn.\eqref{29}--Eqn.\eqref{31}, we need to choose the form of $f(T)$. In the following sections, we will look more closely at three models known to explain some interesting features of the Universe. 

\subsection{Model-I}
We choose the logarithmic form of $f(T)$ \cite{Zhang_2011a, Bamba_2011},
\begin{equation}\label{2A}
f(T) = \beta T \ln\left(\frac{T}{T_{0}}\right) 
 \end{equation}
 
where $\beta$ is an arbitrary model parameter and $T_{0}=-6H_{0}^{2}$ is the present value of torsion scalar $T$. The autonomous system, Eqn. \eqref{29}--Eqn. \eqref{31} become, 
\begin{eqnarray}
\frac{dx}{dN} &=& -\frac{3(x+y-1)(2x+y)}{(2+4\beta -y)} \,,\label{35}\\
 \frac{dy}{dN} &=& -\frac{12\beta  (x+y-1)}{(2+4\beta -y)} \,,\label{36} \\ \nonumber
 \end{eqnarray}
 \begin{eqnarray}
 \frac{d\sigma}{dN}&= -\sigma(\sigma+2) - \frac{3(x+y-1)}{(2-y)} 
 -\frac{3\sigma (x+y-1)}{(2+4\beta -y)} \,. \label{37}
\end{eqnarray}
The corresponding EoS and deceleration parameter reduces to, 
 \begin{eqnarray}
 \omega_{de}&=&\frac{-4 \beta +2 x+y}{(x+y) (-4 \beta +y-2)} \,, \label{38}\\
\omega_{tot}&=& \frac{4 \beta +2 x+y}{-4 \beta +y-2} \label{39} \,, \\
q&=&-1+\frac{3 (x+y-1)}{-4 \beta +y-2} \,.\label{40}
\end{eqnarray}

We can find the critical points by applying the criteria $\frac{dx}{dN}=0$, $\frac{dy}{dN}=0$, and $\frac{d\sigma}{dN}=0$ to the autonomous dynamical system [Eqn.\eqref{35}--Eqn.\eqref{37}]. Four critical points are obtained and has been provided in Table--\ref{TABLE-I} with its corresponding cosmology. In Table--\ref{TABLE-II}, we have derived eigenvalues of the Jacobian matrix. Where  $\lambda_{1}$,  $\lambda_{2}$, and  $\lambda_{3}$ indicate eigenvalues of the Jacobian matrix at both background and perturbation levels.
\begin{widetext}

\begin{table}[H]
    \caption{Critical points of {\bf Model-I}} 
    \centering 
    \begin{tabular}{|c|c|c|c|c|c|c|c|c|c|} 
    \hline\hline 
    C.P. & $x_{c}$ & $y_{c}$ & $\sigma_{c}$  &$\omega_{de}$ &$\omega_{tot}$&$q$&$\Omega_{de}$&$\Omega_{m}$& Exists for \\ [0.5ex] 
    \hline\hline 
    $A_{1}$  & $x$ & $-2x$ & $1$ &$0$&$0$&$\frac{1}{2}$&$-x$&$1+x$ & Always \\
    \hline
    $A_{2}$ &$x$ & $-2x$ & $-\frac{3}{2}$&$0$&$0$&$\frac{1}{2}$&$-x$&$1+x$& Always \\
    \hline
    $A_{3}$ & $x$ & $1-x$ & $-2$  &$-1+\frac{8\beta}{1+4\beta+x}$&$-1$&$-1$&$1$&$0$&Always \\
    \hline
    $A_{4}$ & $x$ & $1-x$ & $0$  &$-1+\frac{8\beta}{1+4\beta+x}$&$-1$&$-1$&$1$&$0$& Always\\
    [1ex] 
    \hline 
    \end{tabular}
    \label{TABLE-I}
\end{table}

\end{widetext}

\begin{table}[H]
    \caption{Eigenvalues and stability condition. } 
    \centering 
    \begin{tabular}{|c|c|c|c|c|} 
    \hline\hline 
    C.P. & Stability Conditions  & $\lambda_{1}$ & $\lambda_{2} $  &$\lambda_{3}$  \\ [0.5ex] 
    \hline\hline 
    $A_{1}$  & Saddle Unstable  & $0$ & $-\frac{5}{2}$ &$3$ \\
    \hline
    $A_{2}$  & Node Unstable  & $0$ & $\frac{5}{2}$ &$3$ \\
    \hline
    $A_{3}$  & Saddle Unstable & $0$ & $-3$ &$2$ \\
    \hline
    $A_{4}$  & Node Stable & $0$ & $-3$ &$-2$\\
    [1ex] 
    \hline 
    \end{tabular}
    \label{TABLE-II}
\end{table}

{\bf Summary of the critical points (Model-I):}
 \begin{itemize}
 \item\textbf{$A_{1}$:}  This critical point represents a matter-dominated scaling solution at the background level. The density parameter for the matter phase is  $\Omega_{m}=1+x$. The total and dark energy sector EoS parameters are respectively $\omega_{tot}=0$, and  $\omega_{de}=0$. At the background level, a positive value of the deceleration parameter $q=\frac{1}{2}$, indicates the decelerated phase of the Universe. We derive $\sigma=1$, at the perturbation level. The positive value of $\sigma$ suggests the growth in matter perturbation. From linear stability theory, both positive and negative eigenvalues indicate unstable saddle behavior. So, this point may be the best way to explain how structures formed when matter dominated, both at the background and perturbation levels.
 
\item\textbf{$A_{2}$:}  At the background level, this critical point is also related to matter dominated phase of the Universe and the both the EoS parameters vanish. The deceleration parameter is, $q=\frac{1}{2}$ shows the decelerating phase of the Universe. Similar to point $A_{1}$, the critical point $A_{2}$ does not exhibit any late time acceleration for any physically accepted value of $q$ and $\omega_{tot}$. Also, we get $\sigma=-\frac{3}{2}$, at the perturbation level, which indicates the decay in matter perturbation. The corresponding eigenvalues of the Jacobian matrix suggest node unstable behavior.

\item\textbf{$A_{3}$:} This critical point is absolutely dark energy dominated solution $\Omega_{de}=1$, with  $\omega_{tot}=-1$ and $q=-1$ at the background level. The negative value of the deceleration parameter shows the accelerating phase of the Universe and $\omega_{tot}=-1$ behaves as a cosmological constant. At the perturbation level, we have derived $\sigma=-2$, which implies the decay in matter perturbation. The eigenvalues of this critical point shows saddle unstable behavior. So, at the perturbation level, this critical point does not exhibit late time acceleration of the Universe.

\item\textbf{$A_{4}$:} This critical point again implies the de-sitter phase solution of the Universe. The corresponding scaling solution of  EoS and deceleration parameters are $\omega_{tot}=-1$ and $q=-1$ respectively at the background level. Similar to critical point $A_{3}$, this point also indicates late time acceleration of the Universe. At the perturbation level, we have accomplished $\sigma=0$. This suggests that the matter perturbation is unchanged. The eigenvalues of this critical point are negative real part and zero. Here, the curve is one dimensional with one vanishing eigenvalue and hence it is hyperbolic \cite{Coley:1999,aulbach1984}. This critical point shows stable node behavior. In addition, this critical point shows the late time acceleration phase of the Universe at both background and perturbation levels. 
\end{itemize}

Two matter-dominated critical points ($A_1, A_2$) and two dark energy-dominated critical points ($A_3, A_4$) are obtained for the considered logarithmic form of $f(T)$. Both the matter-dominated critical points are unstable. The saddle instability, a defined growth rate in a matter perturbation, is represented by critical points $A_1$. The unstable node, represented by critical point $A_2$, represents the decay in matter perturbation. Critical point $A_3$ shows the accelerating behaviour of the Universe only at the background level, whereas the critical point $A_4$ shows the similar behavior both at the background and perturbation levels, however it is stable.     

\begin{figure}[H]
\centering
\includegraphics[width=87mm]{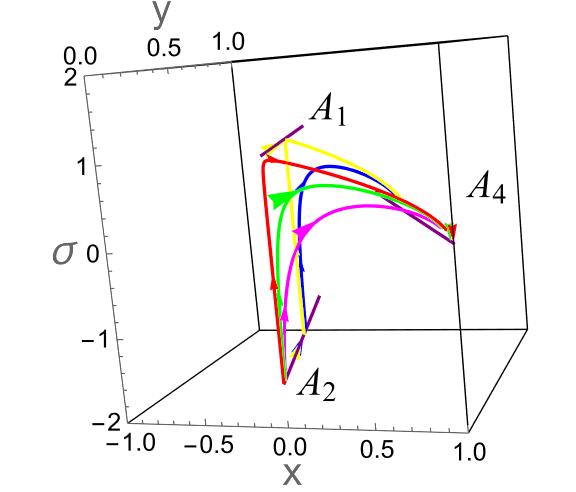}
\caption{3D phase portrait for {\bf Model-I}.} \label{FigF}
\end{figure}

\begin{figure}[H]
\centering
\includegraphics[width=85mm]{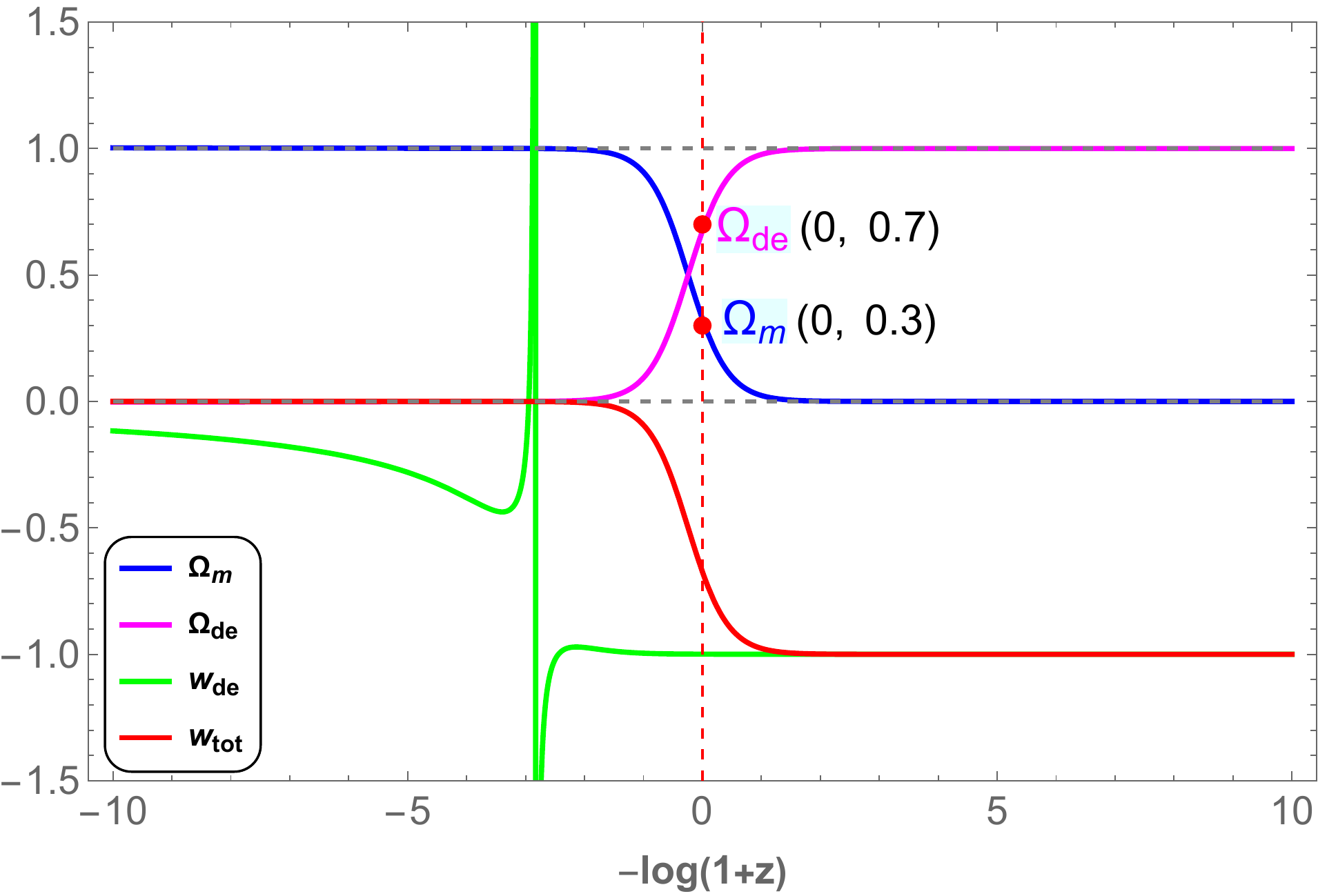}
\includegraphics[width=85mm]{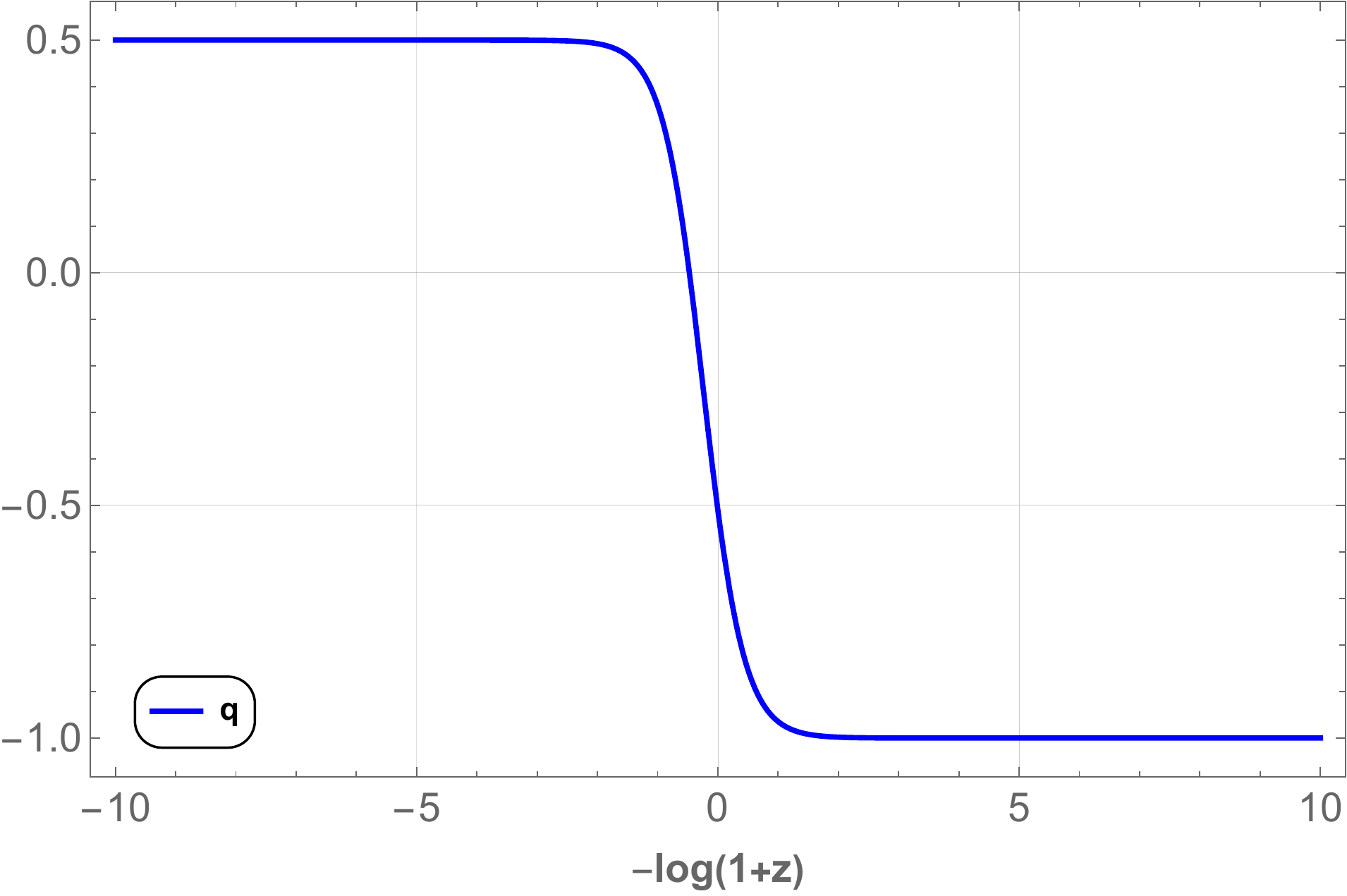}
\caption{Evolution of density parameters (\textbf{Upper panel}) and the deceleration parameter (\textbf{{Lower panel}})  for {\bf Model--I}. The initial conditions $x = 10^{-2}$, $y =10^{-6} $ and $\beta = 0.0001$. The vertical dashed red line denotes the present time.} \label{FigA}
\end{figure}

Fig.--\ref{FigF} displays the phase portrait in $3D$ space. The selected trajectory moved from matter dominated to dark-energy-dominated critical points. From Fig.--\ref{FigF}, we can easily observe the transition of the trajectory like $A_{2}$ (node unstable) \textrightarrow  $A_{1}$ (saddle unstable)  \textrightarrow  $A_{4}$ (node stable). Fig.--\ref{FigA} ({ \bf Upper panel}), represents the evolutionary history of density and the EoS parameters. In the {\bf Upper panel}, the Universe transits from a matter to an acceleration era at late times. The current density parameters for the matter and dark energy sectors are $\Omega_m \approx0.3$ and $\Omega_{de} \approx0.7$, respectively. The total EoS parameter starts with a matter-dominated era ($\omega_{tot}=0$) and approaches the dark-energy sector ($\omega_{tot.}=-1$) at late-time. Also, the dark energy EoS parameter goes to $-1$ at the late phase of the evolution. The present value of the dark energy EoS parameter $\omega_{de}=-1$, matches with the current observational range $\omega_{de}= -1.028 \pm 0.032 $ \cite{Aghanim:2018eyx}. In the { \bf Lower panel}, the deceleration parameter shows transient behavior from deceleration to the acceleration phase of the Universe.  The transition point from deceleration to acceleration is $z=0.59$ and the present value of the deceleration parameter is $q_0=-0.57$ \cite{PhysRevResearch.2.013028a}.

\subsection{Model-II}
We consider the power law form of $f(T)$  \cite{Bengochea:2008gz} as,
\begin{equation}\label{2B}
f(T) = f_{0}(-T)^{m}, 
\end{equation}

where the arbitrary constants $f_{0}$ and $m$ are the model parameters and for this choice of $f(T)$, the dynamical system can be presented as, 
\begin{eqnarray}
\frac{dx}{dN}&=& -\frac{3(x+y-1)(2x+y)}{(2+(1-2m)y)} \,,\label{41}\\
 \frac{dy}{dN}&=& \frac{6y(m-1)(x+y-1)}{(2+(1-2m)y)} \,,\label{42}
 \end{eqnarray}
 \begin{eqnarray}
 \frac{d\sigma}{dN}&= -\sigma(\sigma+2) - \frac{3(x+y-1)}{(2-y)}-\frac{3 \sigma(x+y-1)}{(2+(1-2m)y)} \,.\label{43a}    
\end{eqnarray}
Also, the corresponding EoS and deceleration parameters are,
\begin{eqnarray}
 \omega_{de}&=&\frac{(2 m-1) y+2 x}{((2 m-1) y-2) (x+y)} \,, \label{44}\\
\omega_{tot}&=&-1+ \frac{2 (x+y-1)}{(2 m-1) y-2} \,,\label{45} \\
q&=&-1+\frac{3 (x+y-1)}{(2 m-1) y-2} \,.\label{46}
\end{eqnarray} 

Using the same approach as in {\bf Model--I}, the critical points of the autonomous dynamical system [Eqn. \eqref{41}--Eqn. \eqref{43a}] are summarized in Table--\ref{TABLE-III}. The eigenvalues of the Jacobian matrix are presented in Table--\ref{TABLE-IV}.
\begin{widetext}
 
   \begin{table}[H]
    \caption{Critical points of {\bf Model--II }} 
    \centering 
    \begin{tabular}{|c|c|c|c|c|c|c|c|c|c|} 
    \hline\hline 
    C.P. & $x_{c}$ & $y_{c}$ & $\sigma_{c}$  &$\omega_{de}$ &$\omega_{tot}$&$q$&$\Omega_{de}$&$\Omega_{m}$& Exists for \\ [0.5ex] 
    \hline\hline 
    $B_{1}$  & $0$ & $0$ & $1$ &$-$&$0$&$\frac{1}{2}$&$0$&$1$ & Always \\
    \hline
    $B_{2}$ &$0$ & $0$ & $-\frac{3}{2}$&$-$&$0$&$\frac{1}{2}$&$0$&$1$& Always \\
    \hline
    $B_{3}$ & $x$ & $1-x$ & $-2$  &$\frac{x (2 m-3)-2 m+1}{x (2 m-1)-2 m+3}$&$-1$&$-1$&$1$&$0$&Always \\
    \hline
    $B_{4}$ & $x$ & $1-x$ & $0$  &$\frac{x (2 m-3)-2 m+1}{x (2 m-1)-2 m+3}$&$-1$&$-1$&$1$&$0$& Always\\
    [1ex] 
    \hline 
    \end{tabular}
    \label{TABLE-III}
\end{table}

\end{widetext}

\begin{table}[H]
    \caption{Eigenvalues and stability condition. } 
    \centering 
    \begin{tabular}{|c|c|c|c|c|} 
    \hline\hline 
    C.P. & \begin{tabular}{@{}c@{}}Stability\\ Conditions\end{tabular}    & $\lambda_{1}$ & $\lambda_{2} $  &$\lambda_{3}$  \\ [0.5ex] 
    \hline\hline 
    $B_{1}$  & Unstable  & $3$ & $-\frac{5}{2}$ &$-3(m-1)$ \\
    \hline
    $B_{2}$  &  Unstable  & $3$ & $\frac{5}{2}$ &$-3(m-1)$ \\
    \hline
    $B_{3}$  & Unstable & $0$ & $2$ &$-\frac{3(3-2m+2x-x^{2}+2mx^{2})}{(1+x)(3-2m-x+2mx)}$ \\
    \hline
    $B_{4}$  & \begin{tabular}{@{}c@{}}Stable for\\ $\left(\left.x\right|m\right)\in \mathbb{R}$\end{tabular}    & $0$ & $-2$ &$-\frac{3(3-2m+2x-x^{2}+2mx^{2})}{(1+x)(3-2m-x+2mx)}$\\
    [1ex] 
    \hline 
    \end{tabular}
    \label{TABLE-IV}
\end{table}

{\bf Summary of the critical points (\bf Model--II):}
 \begin{itemize}
\item\textbf{$B_{1}$:} The solution of this critical point represents the matter phase of the Universe. From Table--\ref{TABLE-III},  $\Omega_{m}=1$, the critical point exists for all values of the free model parameter. The values of the total  EoS parameter and deceleration parameter imply that there is no late time acceleration for this solution at the background level. The value of $\omega_{de}$ is undefined for this critical point.  At the perturbation level, we find $\sigma=1$, and the positive value of $\sigma$  implies a growth factor in matter perturbation. This critical point shows saddle unstable behavior.  

\item\textbf{$B_{2}$:} At the background level, this critical point is similar to critical point $B_{1}$. We have discovered $\sigma=-\frac{3}{2}$ at the perturbation level, which denotes a decline in matter perturbation. This critical point exhibits node unstable behavior for ($1>m$) and saddle unstable for ($m>1$).    \\

\item\textbf{$B_{3}$:} This critical point is having dark energy-dominated solution with $\Omega_{de}=1$. With $\omega_{tot}=-1$, and $q=-1$, the values of EoS and deceleration parameter shows late time acceleration of the Universe at the background level. But at the perturbation level, we find $\sigma=-2$. The negative value of perturb variable indicates decay in matter perturbation. The eigenvalues of the Jacobian matrix imply unstable behavior for any value of $x$ and $m$. However, this critical point does not exhibit accelerating behavior at the perturbation level. It only indicates acceleration at the background level. 

\item\textbf{$B_{4}$:} At the background level both $B_{3}$  and $B_4$ are having similar behaviour. At the perturbation level, we obtain $\sigma=0$, which indicates that the perturbation of matter is unchanged. It shows the stable behavior for any choice of $m$ and $x$ \cite{Coley:1999,aulbach1984} and exhibits late time acceleration at both perturbation and background level. 
 \end{itemize} 

For the power law form of $f(T)$, two matter-dominated critical points ($B_1, B_2$) and two dark energy-dominated critical points ($B_3, B_4$) are obtained. Among these, one stable critical point $B_{4}$, which represents accelerated expansion and dark energy-dominated phase of the Universe. The second dark energy-dominated critical point $B_{3}$ shows saddle instability, decay in matter perturbation, and accelerated expansion of the Universe only at the background level. The two unstable matter-dominated critical points, $B_{1}$ and $B_{2}$ show decelerated phase of the Universe.   

 \begin{figure}[H]
\centering
\includegraphics[width=82mm]{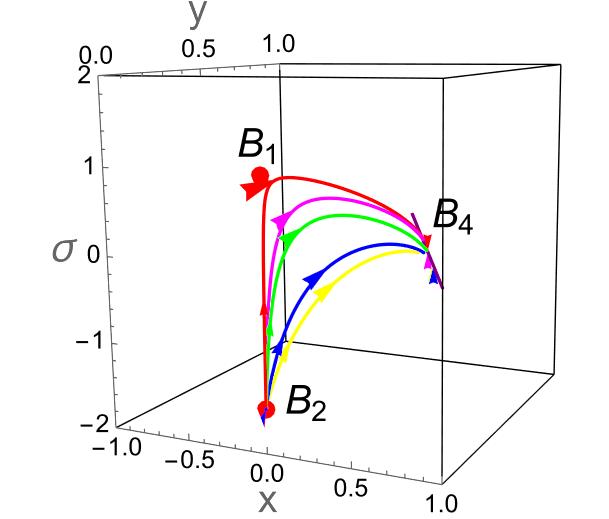}
\caption{3D phase portrait for {\bf Model-II}.} \label{FigE}
\end{figure}

\begin{figure}[H]
\centering
\includegraphics[width=88mm]{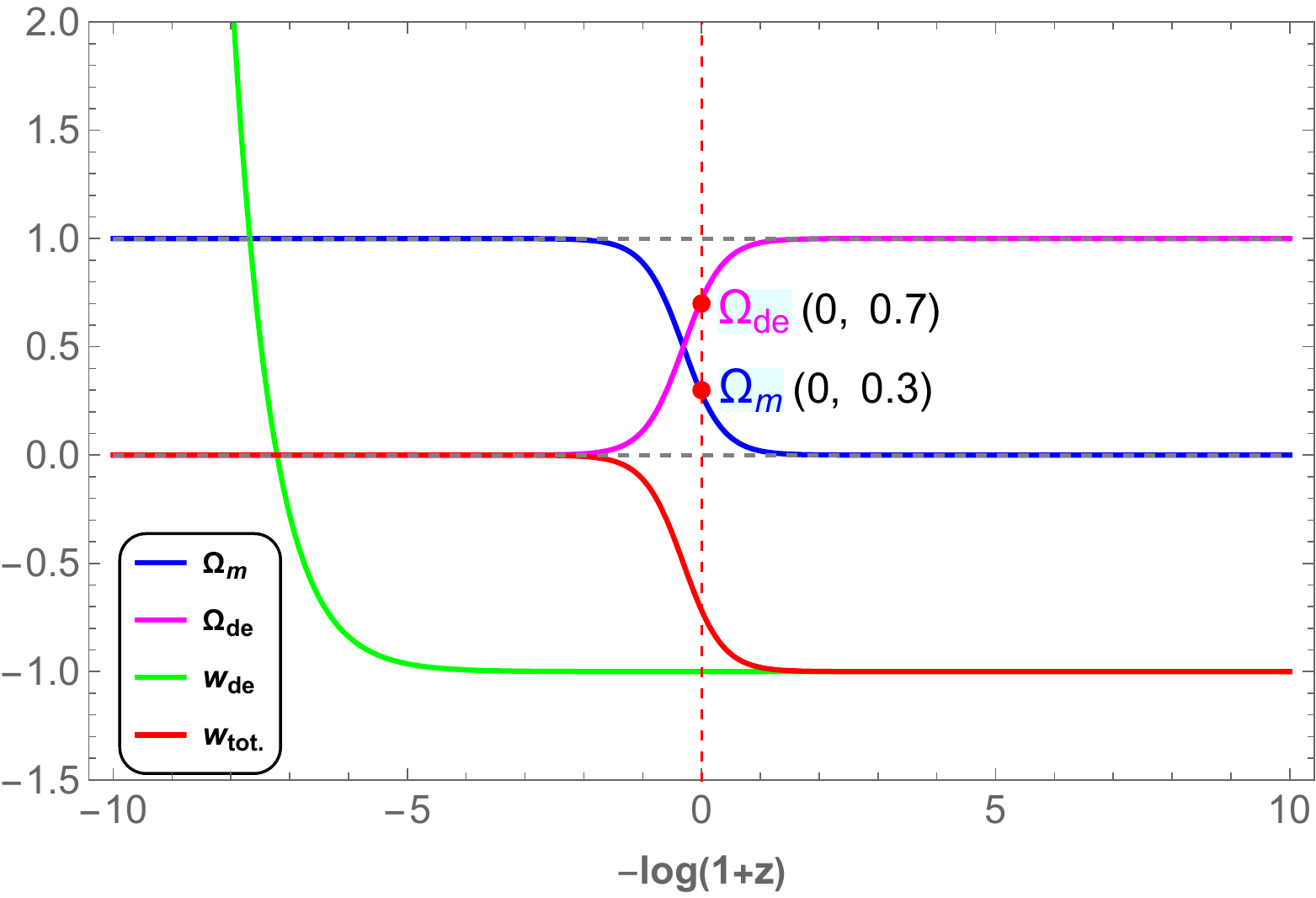}
\includegraphics[width=88mm]{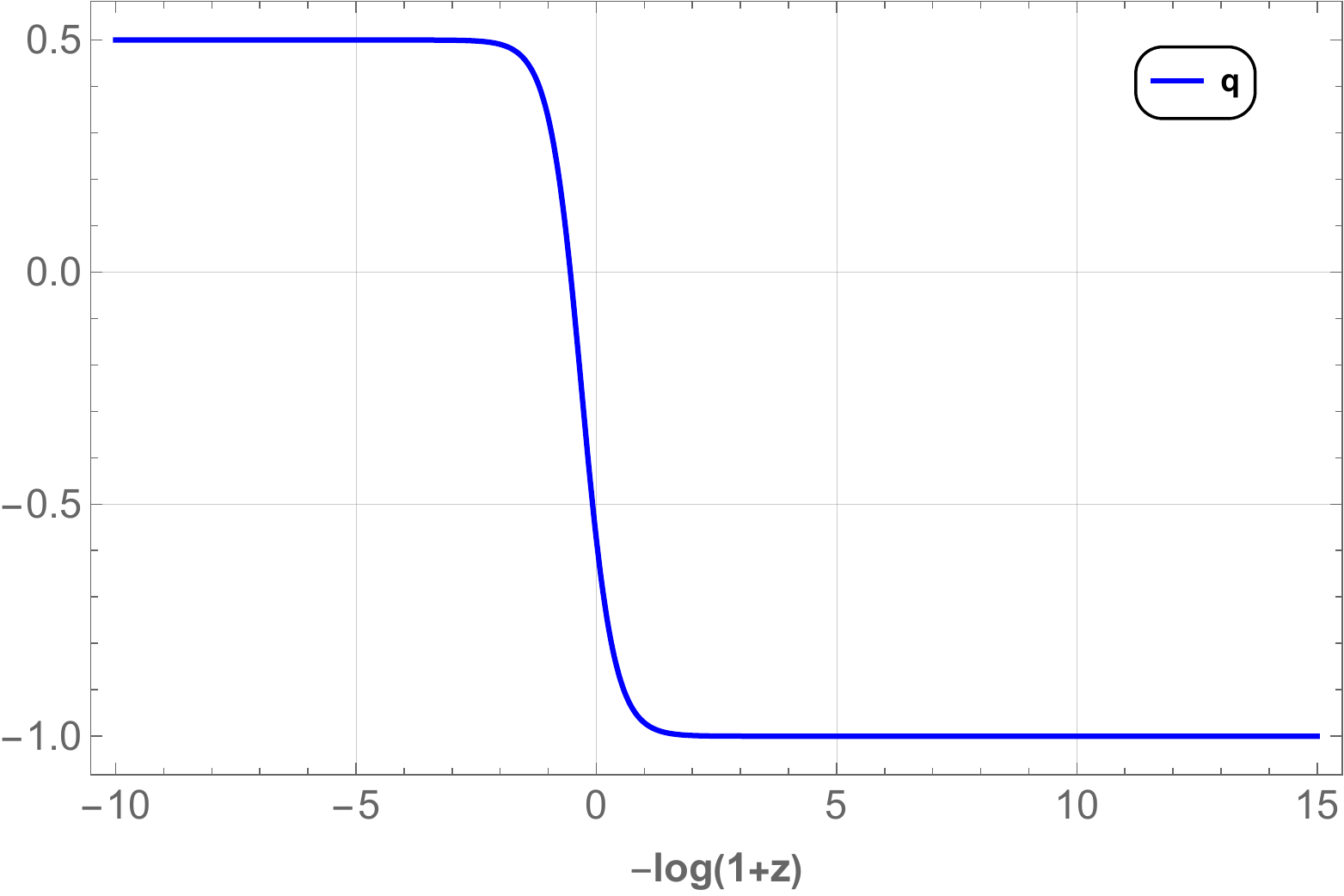}
\caption{Evolution of density parameters (\textbf{Upper panel}) and the deceleration parameter (\textbf{\textbf{Lower panel}})  for {\bf Model--II}. The initial conditions $x = 10^{-3}$, $y =10^{-6} $ and $m = 0.5$. The vertical dashed red line denotes the present time.} \label{FigC}
 \end{figure}

Fig.--\ref{FigE} shows the critical points in 3D space. The trajectories show a path from the matter-dominated unstable critical points $B_{1}$ and $B_{2}$ to the stable dark energy-dominated critical point $B_{4}$ ($B_{2}$\textrightarrow $B_{1}$\textrightarrow  $B_{4}$). Fig.--\ref{FigC} ({\bf Upper panel}) shows the evolution of the density and EoS parameters with respect to redshift $z$. The initial condition values are calibrated to eventually give present time (at $z=0$) values of both density parameters. The graph is in accordance with the expected behavior of the Universe to transition from a matter-dominated phase to a dark energy-dominated phase. Consequently, Fig.--\ref{FigC} ({\bf Lower panel}) shows transitions from decelerated to accelerated expansion at $z=0.64$ and the present value of the deceleration parameter is noted as $q_{0}=-0.62$ \cite{PhysRevResearch.2.013028a}. As shown in the {\bf Upper panel}, the value of dark energy EoS parameter $\omega_{de}=-1$ is also in accordance with the current observational value \cite{Aghanim:2018eyx}.

\subsection{Model-III}
Finally, we consider the form of $f(T)$ as the combined form of logarithmic and power law  \cite{Mirza_2017a}, 
\begin{equation}\label{42}
 f(T)=\alpha (-T)^{n}  \ln\left(\frac{T}{T_{0}}\right),   
\end{equation}
where $\alpha$ and $n$ are arbitrary constants and for this $f(T)$, the autonomous system becomes, 
\begin{eqnarray}
\frac{dx}{dN}&=&-\frac{3 (x+y-1) (2 x+y)}{-4 n (n x+y)+y+2} \,,\label{43}\\
\frac{dy}{dN}&=&-\frac{6 (x+y-1) (y-2 n (n x+y))}{-4 n (n x+y)+y+2} \,,\label{44}
\end{eqnarray}
\begin{eqnarray}
\frac{d\sigma}{dN}&=-\frac{3 \sigma  (x+y-1)}{-4 n (n x+y)+y+2}-\sigma  (\sigma +2)+\frac{3 (x+y-1)}{y-2} \,.\label{45}
\end{eqnarray}
and the EoS parameters and deceleration parameter could be,
\begin{eqnarray}
\omega_{de} &=& \frac{y-2 \left(2 n^2 x+2 n y+x\right)}{(x+y) (-4 n (n x+y)+y+2)} \,,\label{46}\\
\omega_{tot} &=&-1 -\frac{2 (x+y-1)}{-4 n (n x+y)+y+2} \,, \label{47}\\
q &=&-1 -\frac{3 (x+y-1)}{-4 n (n x+y)+y+2} \,.\label{48}
\end{eqnarray}

In a similar approach, the critical points of the autonomous dynamical system represented by [Eqn.\eqref{43}--Eqn.\eqref{45}] are calculated and are presented in Table-\ref{TABLE-V}. The corresponding eigenvalues of the Jacobian matrix are listed in Table-\ref{TABLE-VI}.
\begin{widetext}

   \begin{table}[H]
\caption{Critical points of {\bf Model--III} } 
\centering 
\begin{tabular}{|c|c|c|c|c|c|c|c|c|c|} 
\hline\hline 
C.P. & $x_{c}$ & $y_{c}$ & $\sigma_{c}$  &$\omega_{de}$ &$\omega_{tot}$&$q$&$\Omega_{de}$&$\Omega_{m}$& Exists for \\ [0.5ex] 
\hline\hline 
\hline
$C_{1}$ & $0$ & $0$ & $1$  &$-$&$0$&$\frac{1}{2}$&$0$&$1$&Always \\
\hline
$C_{2}$ & $0$ & $0$ & $-\frac{3}{2}$  &$-$&$0$&$\frac{1}{2}$&$0$&$1$& Always\\
\hline
$C_{3}$ &$x$ & $1-x$ & $-2$&$\frac{4n-1+x(3+4n(n-1)}{4n-3+x(1-2n)^{2}}$&$-1$&$-1$&$1$&$0$& Always \\
\hline
$C_{4}$  & $x$ & $1-x$ & $0$ &$\frac{4n-1+x(3+4n(n-1)}{4n-3+x(1-2n)^{2}}$&$-1$&$-1$&$1$&$0$ & Always \\
[1ex] 
\hline 
\end{tabular}
\label{TABLE-V}
\end{table}

\end{widetext}

\begin{table}[H]
\caption{Eigenvalues and stability condition. } 
\centering 
\begin{tabular}{|c|c|c|c|c|} 
\hline\hline 
C.P. &  \begin{tabular}{@{}c@{}}Stability\\ Conditions\end{tabular}   & $\lambda_{1}$ & $\lambda_{2} $  &$\lambda_{3}$  \\ [0.5ex] 
\hline\hline 
\hline
$C_{1}$  & Unstable & $-\frac{5}{2}$ & $-3(n-1)$ &$-3(n-1)$ \\
\hline
$C_{2}$  & Unstable & $\frac{5}{2}$ & $-3(n-1)$ &$-3(n-1)$\\
\hline
$C_{3}$  &  Saddle Unstable  & $0$ & $-3$ &$2$ \\
\hline
$C_{4}$  & Stable   & $0$ & $-3$ &$-2$ \\
[1ex] 
\hline 
\end{tabular}
\label{TABLE-VI}
\end{table}
{\bf Summary of the critical points (Model-III):}
\begin{itemize} 
\item\textbf{$C_{1}$:} The eigenvalues of this critical point show saddle unstable behavior for ($1>n$). The value of dark energy EoS parameter ($\omega_{de}$) is undefined and total EoS parameter is, $\omega_{tot}=0$. Also, the deceleration parameter, $q=\frac{1}{2}$ which indicates decelerated expansion of the Universe at the background level. The positive value of $\sigma=1$ shows increase in matter perturbation and consequently non-decelerated expansion of the Universe at perturbation level. This critical point resembles matter dominated phase as $\Omega_{m}=1$.

\item\textbf{$C_{2}$:} This critical point shows similar behaviour as that of $C_{1}$ except the value of $\sigma$. Here, we obtain, $\sigma=-\frac{3}{2}$ which suggests a decay in matter perturbation and decelerated expansion of the Universe at both background and perturbation levels. This critical shows saddle unstable behavior for ($n>1$) and node unstable for ($1>n$). 

\item\textbf{$C_{3}$:} This critical point indicates dark energy dominated solution with $\Omega_{de}=1$ and $\Omega_{m}=0$. Since, the values of deceleration parameter and total EoS parameter are respectively, $q=-1$ and $\omega_{tot}=-1$. This critical point shows accelerated expansion of the Universe at the background level, but not at the perturbation level, as $\sigma = -2$ which resembles decay of matter perturbation. The eigenvalues show saddle unstable behaviour as the sign of non zero eigenvalues is positive.

\item\textbf{$C_{4}$:} This is the only critical point that shows stable behaviour because the non zero eigenvalues are negative. Since $\sigma = 0$, we observe that the matter perturbation remains constant. This also resembles entirely dark energy dominated phase and late time accelerated expansion of the Universe at both, background and perturbation levels.
\end{itemize}

For this combined form of $f(T)$, we examine two matter-dominated critical points ($C_1, C_2$) and two dark energy-dominated critical points ($C_3, C_4$) like the other two models. Both the matter-dominated critical points $C_1$ and $C_2$ are unstable and indicate decelerated phase of the Universe. One of the dark energy-dominated critical points $C_3$ shows saddle instability and accelerated expansion of the Universe at the background level but not at the perturbation level. The other dark energy-dominated critical point $C_4$ is stable, showing the accelerated expansion of the Universe at both levels and no change in matter perturbation.

\begin{figure}[H]
\centering
\includegraphics[width=85mm]{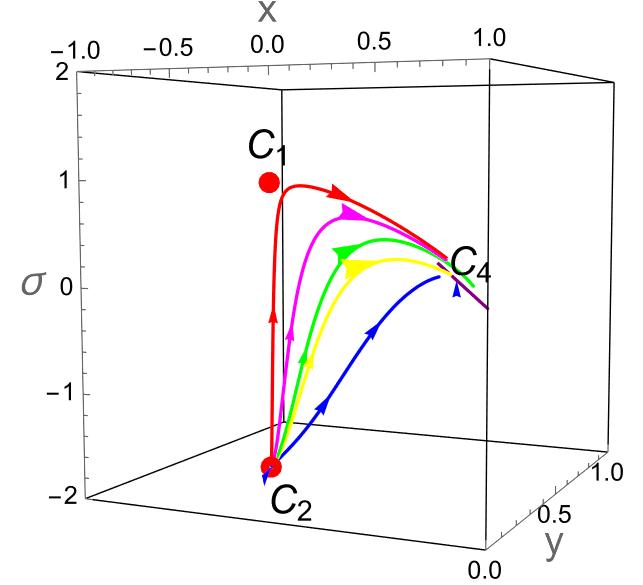}
\caption{3D phase portrait for {\bf Model-III}.} \label{FigG}
\end{figure}
 \begin{figure}[H]
\centering
\includegraphics[width=88mm]{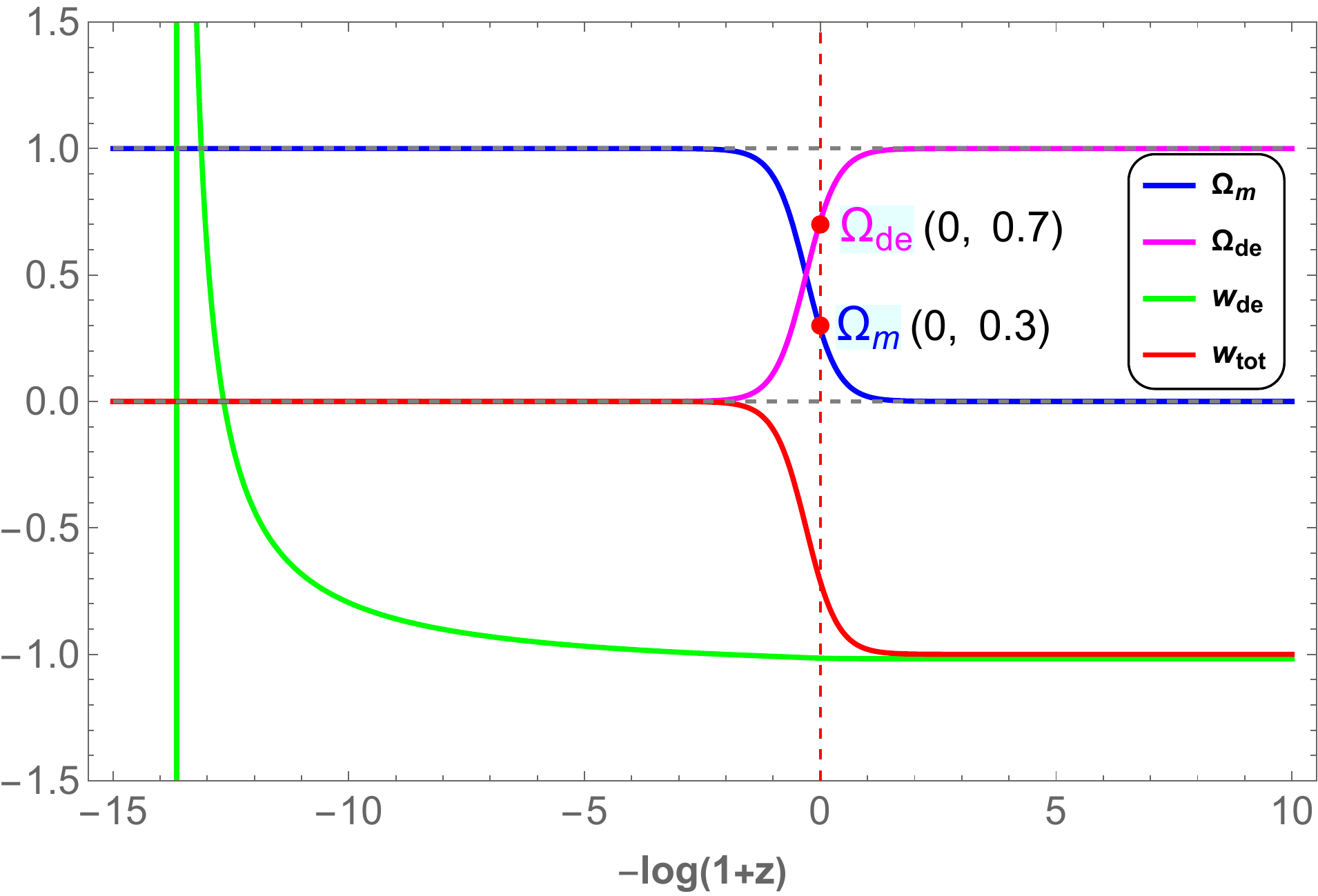}
\includegraphics[width=88mm]{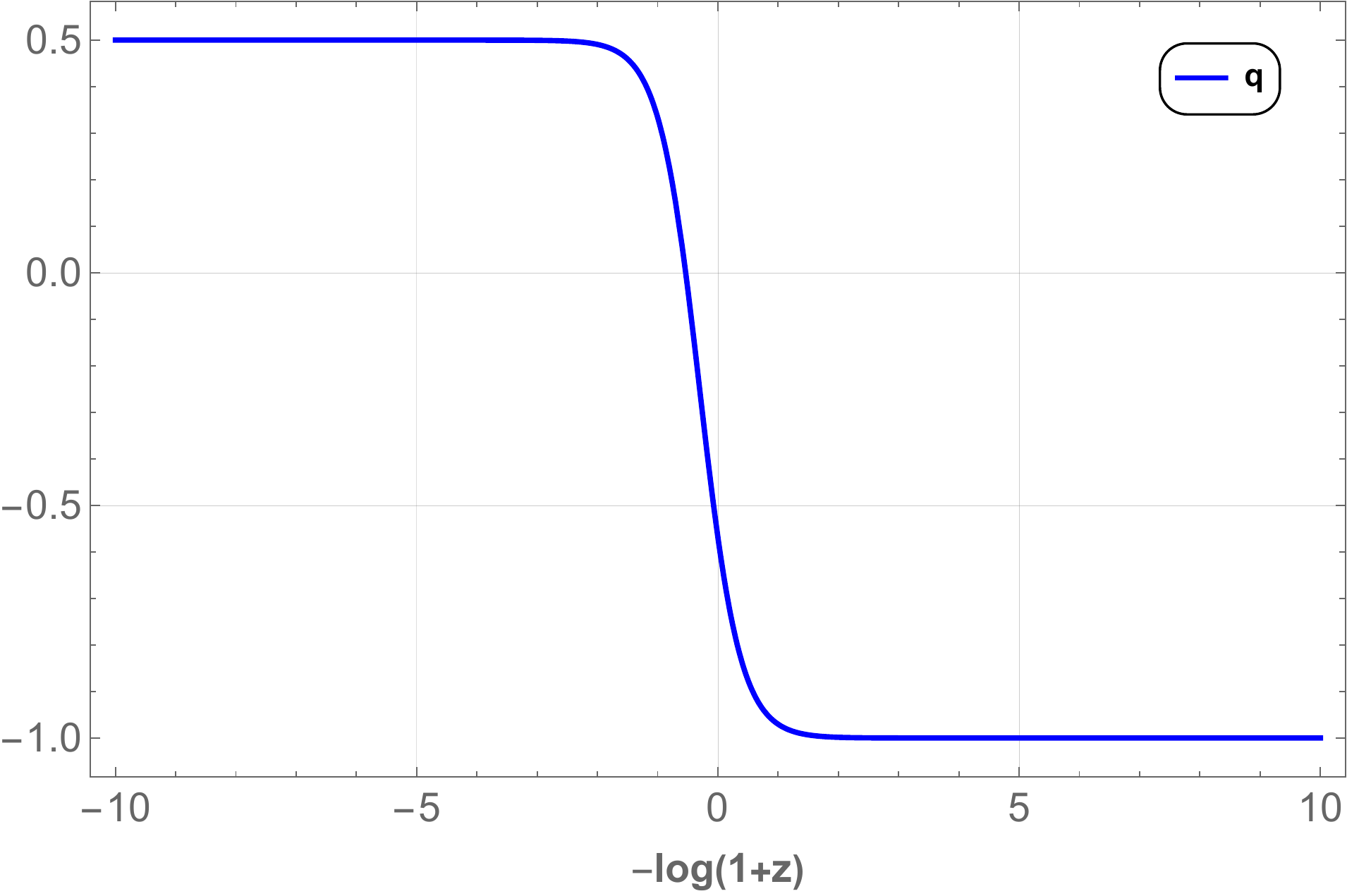}
\caption{Evolution of density parameters (\textbf{Upper panel}) and the deceleration parameter (\textbf{\textbf{Lower panel}})  for {\bf Model--III}. The initial conditions $x = 10^{-2}$, $y =10^{-6} $ and $n = 0.03$. The vertical dashed red line denotes the present time.} \label{FigH}
 \end{figure}

Fig.--\ref{FigG} shows the critical points and the trajectories, $C_{2}$\textrightarrow $C_{1}$\textrightarrow  $C_{4}$ in $3D$ space. The trajectories show paths from matter to dark energy-dominated critical points. Fig.--\ref{FigH} shows the evolution of all the background parameters with respect to redshift. It is in accordance with our observation that matter-dominated critical points represent decelerated phase and dark energy-dominated critical points represent the accelerated expansion phase of the Universe. The initial condition values are set such that they give present time values of both, matter and dark energy density parameters as shown in Fig.--\ref{FigH} ({\bf Upper panel}). It also shows that both dark energy and total EoS parameters go to $-1$ at the late time phase of the Universe. Fig.--\ref{FigH} ({\bf Lower panel}) shows a transition from decelerated to the accelerated expansion of the Universe at redshift $z=0.68$. The present value of the deceleration parameter obtained as, $q_{0}=-0.56$ \cite{PhysRevResearch.2.013028a}.

\section{Conclusion} \label{SEC-IV}
Dynamical system analysis is an useful approach to analyse the qualitative behavior of the Universe. In this approach, we deal with the non-linear differential equation in terms of dynamical variables. This concept describes the evolution of the Universe through the critical points of autonomous systems. Taking these things into consideration, we examined dynamical system analysis in $f(T)$ gravity both at the background and perturbation levels in this study. In this approach, we have described general dynamical autonomous systems [Eqn.\eqref{29}-- Eqn.\eqref{31}] in the teleparallel framework, to be specific in $f(T)$ gravity. The dynamical variable $x$ and $y$ represents the background behavior of the Universe whereas the dynamical variable $\sigma$ defined the perturbed level of the Universe i.e the growth and decay in matter perturbation. In the defined autonomous systems, the functional form of $f(T)$ are incorporated; three distinct forms of $f(T)$ are proposed leads to three different models.

In {\bf Model--I}, we have taken the logarithmic form of $f(T)$, which is displayed in Eqn.\eqref{2A}. For this form of $f(T)$, we have obtained four critical points, which are defined matter and the dark-energy era of the Universe at both background and perturbation levels. The critical points $A_{1}$ and $A_{2}$ describe the matter-dominated era, but  $A_{1}$ and $A_{2}$ show a growth rate and decay in matter perturbation respectively. The critical points $A_{3}$ and $A_{4}$ represent the dark-energy era of the Universe, but at the background level, the critical point $A_{3}$ shows acceleration expansion of the Universe and in perturbation level shows the decay in matter perturbation. But, the critical point $A_{4}$ shows accelerated expansion of the Universe at both levels. Also, it is showing stable node behavior. In {\bf Model--II}, we have considered the power-law form of $f(T)$, which is presented in Eqn.\eqref{2B}, and four critical points have been obtained. The behavior of critical points for this model is similar to that of {\bf Model--I} in spite of different form of $f(T)$. critical points $B_{1}$ and $B_{2}$ determined the matter phase of the Universe and critical points $B_{3}$ and $B_{4}$ show the dark energy phase of the Universe but here, only critical point $B_{4}$ describes late time acceleration of the Universe at both levels. When we combine both the models in another model as in Eqn.\eqref{42}, the same four critical points are obtained. The qualitative behavior of this model is similar to the first two models both at the background and perturbation levels. We wish to mention here that the cosmological perturbation has been studied to check the stability of the cosmological models in $f(T)$ gravity in Refs. \cite{Coley_2307_12930, Coley_2310_14378}. The investigation was basically on the class of Einstein teleparallel geometries that would have four dimensional Lie algebra of affine connection. Further the explicit form of $f(T)$ has been obtained for each parameter value. We have considered three such forms of $f(T)$ to show the late time cosmic acceleration of the Universe through the dynamical system analysis.. 

The cosmological behaviour of the Universe through the density parameters of matter and dark energy, EoS parameters, and deceleration parameters are shown. From the behaviour of deceleration parameter, it has been observed that in all models, the Universe shows early deceleration to late time acceleration with the transition noted respectively as:  $z=0.59$, $z=0.64$, and $z=0.68$. Also the present value are: $q_0=-0.57$, $q_0=-0.62$, and $q_0=-0.56$. All three models have the same dark EoS parameter value at present time, which is $\omega_{de}=-1$. In all three models, the density parameters for matter and dark energy are $\Omega_{m} \approx 0.3$ and $\Omega_{de} \approx 0.7$, which is consistent with recent observations in cosmology. In 3D space, we have drawn phase space trajectories for all models. The trajectory moves from a matter-dominated (unstable) to a dark energy-dominated (stable) phase. Finally, we conclude that the dynamical stability analysis can be used extensively to analyse the cosmological behaviour of the Universe.

\section*{Acknowledgements}
LKD acknowledges the financial support provided by University Grants Commission (UGC) through Senior Research Fellowship UGC Ref. No.: 191620180688 to carry out the research work. BM acknowledges the support of IUCAA, Pune (India) through the visiting associateship program.

\section{References} \label{SEC-VI} 

\bibliographystyle{utphys}
\bibliography{references}
\end{document}